\documentclass[twocolumn,prl,aps,showpacs,superscriptaddress,floatfix]{revtex4}
\usepackage{epsfig}
\newcommand{\fig}[1]    {Fig.~\ref{#1}}
\newcommand{\subbox}[1] {{\mbox{\scriptsize #1}}}

\def\etal       {{\em et al.~}}

\def\bi         {\begin{itemize}}
\def\ei         {\end{itemize}}
\def\benu       {\begin{enumerate}}
\def\eenu       {\end{enumerate}}
\def\bmat       {\left[ \begin{array}}
\def\emat       {\end{array} \right]}
\def\beq        {\begin{equation}}
\def\eeq        {\end{equation}}
\def\beqn       {\begin{eqnarray*}}
\def\eeqn       {\end{eqnarray*}}
\def\beqa       {\begin{eqnarray}}
\def\eeqa       {\end{eqnarray}}
\def\bquote     {\begin{quote}}
\def\equote     {\end{quote}}
\def\f          {\frac}
\def\bwide      {\begin{widetext}}
\def\ewide      {\end{widetext}}

\def\a          {\alpha}
\def\b          {\beta}
\def\d          {\delta}

\def\k          {\kappa}

\def\m          {\mu}

\def\bk         {{\bf k}}

\def\bq         {{\bf q}}

\def\hx         {{\hat{x}}}

\def\vi         {{\vec{\imath}}}

\def\dag        {\dagger}

\begin{document}

%%%%%%%%%%%%%%%% Header information %%%%%%%%%%%%%%%%%
\title{Nematic domains and resistivity in
an itinerant metamagnet
%the itinerant metamagnet of Sr$_3$Ru$_2$O$_7$
coupled to a lattice}
\author{Hyeonjin Doh}
%\email{hdoh@physics.utoronto.ca}
\affiliation{Department of Physics, University of Toronto, Toronto, 
Ontario M5S 1A7, Canada}
\author{Yong Baek Kim}
%\email{ybkim@physics.utoronto.ca}
\affiliation{Department of Physics, University of Toronto, Toronto, 
Ontario M5S 1A7, Canada}
\author{K. H. Ahn}
\affiliation{Department of Physics, New Jersey Institute of Technology, Newark, NJ 07102 USA}
%\email{kenahn@aps.anl.gov}

\date{\today}

\begin{abstract}
The nature of the emergent phase near a putative quantum critical point
in the bilayer ruthenate Sr$_3$Ru$_2$O$_7$ has been a recent subject of 
intensive research. It has been suggested that this phase may possess
electronic nematic order(ENO). In this work, we investigate the possibility of nematic 
domain formation in the emergent phase, using a phenomenological
model of electrons with ENO and its coupling to lattice degrees of freedom. 
The resistivity due to the scattering off the domain walls is shown to 
closely follow the ENO parameter. Our results provide qualitative 
explanations for the dependence of the resistivity on external magnetic fields
in  Sr$_3$Ru$_2$O$_7$.
\end{abstract}
\pacs{71.27.+a,75.30.Kz,71.10.-w,63.20.Kr}

\maketitle
%%%%%%%%%%%%%% End of Header %%%%%%%%%%%%%%%

%\section{Introduction}

Quantum critical behavior in metamagnetism
represents the most unusual critical phenomenon in itinerant electron systems. 
It was shown in a series of remarkable experiments that the critical end point 
associated with the metamagnetic transition(MMT)
can be driven to zero temperature 
in the bilayer ruthenate Sr$_3$Ru$_2$O$_7$ by 
changing the angle between the external magnetic field and the $c$-axis of the material.\cite{Grigera01Science,perry01}
The transport and thermodynamic measurements reveal unusual temperature 
dependence near the putative {\it quantum} critical end point,
indicating that the underlying itinerant electrons are strongly interacting with
the quantum critical fluctuations of the 
magnetization.\cite{Grigera01Science,perry01,Millis02prl,Green05prl}

Later experiments on the ultra-pure samples, however, 
have shown that the quantum critical end point is eventually avoided when the finite 
temperature critical point is pushed further down to zero temperature,
and is replaced by two consecutive
%metamagnetic transitions.\cite{Grigera04Science,perry04,kitagawa} 
MMTs.\cite{Grigera04Science,perry04,kitagawa} 
The phase bounded by two
MMTs exhibits unusually high resistivity
and its nature has been a recent subject of intensive research.

Currently there exist two competing proposals for the nature of this {\it emergent} phase.
%Binz and collaborators
Binz \etal\cite{Binz06prl}
proposed a phenomenological Ginzburg-Landau theory 
of magnetic Condon domains and suggested that the high resistivity may
come from the scattering of electrons off the magnetic domains.
Another proposal involves the formation of ENO in the emergent
phase; namely the Fermi surface of the electrons spontaneously breaks the 
lattice rotational symmetry.\cite{Grigera04Science,KeeHY05prb} 
Two possible orientations of the ENO may lead to 
domain formation and may indeed lead to high resistivity due to domain
wall scattering.

Very recently it has been reported that the high resistivity seen in the 
emergent phase is insensitive to demagnetization factor or shapes of
the samples.\cite{Borzi07Science} 
This is at odds with a key prediction of the magnetic Condon
domain theory; the system is supposed to be very sensitive to
demagnetization factor.\cite{Binz06prl} 
This may leave only the theory of ENO as an alternative proposal.

In this paper, we offer a theory of the nematic domain formation
and the resulting high resistivity in the emergent phase. 
The possibility of the nematic domains has not been 
studied theoretically even though such possibility was briefly discussed 
in the literature.\cite{Grigera04Science,KeeHY05prb}  
Our objective in the current work is to provide a 
minimal theoretical description that can explain key experimental observations 
at qualitatively level. 

In search for a concise description, we start with an effective electronic 
Hamiltonian that captures the essential physics of ENO, 
namely we consider the quadrupolar interaction between 
electrons.\cite{KeeHY05prb,Oganesyan01prb}
This interaction can be regarded as the angular-momentum-two 
channel of some general interaction or the Landau quasiparticle interaction.
In real materials, the interaction in all angular momentum channels may exist, 
not just in the quadrupolar channel. The dominance of the quadrupolar
channel is an assumption of our phenomenological theory and
its validity should be tested by comparing the results with the experimental
data. While it has not been proven, it is
conceivable that such an instability or a dominant interaction
is more likely to happen near an itinerant quantum critical point
because the Fermi surface would be very soft due to strong 
critical fluctuations.

In view of the magnetostriction data indicating a close relationship
between the lattice and
%the formation of
the emergent 
phase \cite{Grigera04Science}, 
we also consider the electron-lattice interaction. 
Since we are studying an effective Hamiltonian, we are only
interested in general properties of this Hamiltonian and
the precise values of the parameters should not be taken seriously.
Remarkably this simplified model contains essential ingredients 
of the nematic domain formation and the resulting high
resistivity. It is found that the electron-lattice interaction 
makes the formation of the nematic domains (at finite temperatures)
much more likely by offering a large number of metastable
%nematic 
domain configurations. Another key result of our work is that 
the magnetic field dependence of the resistivity in the nematic 
domain phase closely follows the field dependence of the 
ENO parameter. Thus the resistivity can be regarded
as a measurement of the ENO.

{\bf Mean field theory of the nematic order}:
The simplest model for the ENO on the square lattice
can be written by keeping only the interactions in the quadrupolar
channel\cite{Oganesyan01prb}  or the lattice
equivalent of the angular momentum $l=2$ channel in the
continuum\cite{Khavkine04prb}.
\beq
H_\subbox{el}
= \sum_\bk \xi_\bk c_\bk^\dag c_\bk^{}  
+\!\!\!\sum_{\bk, {\bf k'}, \bq,\a} \!\!\!
F (\bq)
%f (\bk,\bk')
f_{\bk\alpha}f_{\bk\alpha}
c_{\bk+\bq}^\dag c_{\bk'\!-\bq}^\dag
c_{\bf k'} c_{\bk}
\label{eqn:nematic-hamil}
\eeq
where $\xi_\bk = -t (\cos k_x + \cos k_y) -2t' \cos k_x \cos k_y-\m$ is
the single-particle dispersion and $\a=1,2$. 
$f_{\bk1} = \cos k_x - \cos k_y$ and 
$f_{\bk2} = \sin k_x \sin k_y$ are the form factors in the 
quadrupolar channels. 
$F({\bf q})$ represents a short range
interaction in real space with $F({\bf q}\!\rightarrow\!0)=F={\rm constant}$.
%$F(\bq) = F$, where $F$ is a constant.
%we take
%$F_\a({\bf q}) = F_\a/(1 + \kappa q^2)$. 
%$F_\a({\bf q}) = F_\a$. 
Here we suppress the spin indices and it should be understood 
that the interaction terms exist in both spin channels with possibly
different interaction strength.

The ENO parameter is
%given by
%\beq
$Q(\bq) = F \sum_\bk f_{\bk1} \langle c_{\bk+\bq}^\dag c_{\bk}^{} \rangle$.
%\eeq
In the unform state $\bq\rightarrow 0$, this order parameter
leads to the renormalized dispersion
in the mean field theory, 
$\xi^{\rm ren}_{\bf k} = -(t + Q(0)) \cos k_x - (t-Q(0)) \cos k_y
-2t'\cos k_x \cos k_y-\m$.
Clearly, the resulting Fermi surface would break
%the $x$-$y$ or 
the $\pi/2$ rotation symmetry of the lattice.
An alternative ENO parameter with $f_{\bk2}$ 
vanishes for the interaction form chosen here.\cite{Khavkine04prb}
%The alternative order
%parameter with $f_2({\bf k})$ form factor is always zero
%for $F_1\ge F_2$ 
%so that it will not be considered here.\cite{Khavkine04prb}

In the presence of a magnetic field, the Zeeman 
energy leads to different chemical
potentials for the spin-up and -down electrons, $\mu_{\uparrow, \downarrow}
= \mu \pm \mu_B H$, where $\mu_B$ is the Bohr magneton and 
$H$ is the external magnetic field. As the magnetic field increases, 
the Fermi surface volume of the majority (minority) spin 
increases (decreases). When the Fermi surface of the majority or
minority spin touches the van-Hove singularity, a first order 
phase transition to a nematic phase occurs with
a sudden opening and elongation of Fermi surface
along $a$- or $b$-axis.\cite{KeeHY05prb}
%in the nematic order parameter.
% When it happens, the Fermi
%surface of the majority or minority spin suddenly changes
%to an open Fermi surface elongated along the $a$- or $b$- axis.
Here the Fermi surface volume of the majority or minority spin jumps,
leading to an MMT.
When the magnetic field increases further, the other side of Fermi surface 
touches the van-Hove singularity again. This leads to another first
order transition 
%where the Fermi surface volume of the majority 
%or minority spin jumps 
and the lattice rotational symmetry is restored afterwards.\cite{KeeHY05prb}

In the uniform nematic state, the spontaneously
chosen nematic direction can be either the $a$- or $b$- axis.
On the other hand,
non-uniform solutions
can arise as meta-stable states in the mean field theory,
In particular,
the nematic domains with different orientations can coexist. 
These static domain structures, however, may not be favorable even 
as meta-stable states in two and one dimensions because of large 
quantum fluctuations. Thus if the nematic domains were to arise,
there should be additional degrees of freedom.

{\bf Normal modes of lattice distortions}:
Magnetostriction effect observed in Sr$_3$Ru$_2$O$_7$ suggests
that the electron-lattice interaction is a natural degree of freedom 
to consider. In the uniform nematic-ordered state, 
%the rotational symmetry of the lattice is broken and 
the electron-lattice interaction
would induce a tetragonal lattice distortion.
This system-wide shape change, however,
is energetically very costly and the system may rather prefer to have
twin boundaries between the phases with different nematic orientations
so that the overall shape is preserved. Thus the 
electron-lattice interaction may indeed prefer the formation of
the nematic domains.

\begin{figure}[htb]
\epsfxsize=7cm
\epsffile{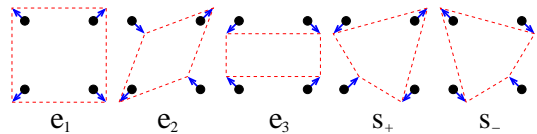}
\caption{Normal modes \label{fig:NormalModes} }
\end{figure}
We treat the lattice classically assuming slow lattice
dynamics.
%We assume that the lattice dynamics is relatively slow so that one
%can treat the lattice classically.
For simplicity, we will only 
consider two-dimensional lattice distortions even though the full 
analysis of the electron-lattice coupling may
require a three dimensional lattice structure. 
%Thus we will consider the normal
%modes of the lattice distortions only inside the two dimensional plane.
Implications to the distortions out of plane will be discussed later.

It is useful to write the lattice Hamiltonian in terms of distortion modes
of a lattice plaquette. 
%There are eight independent modes because 
%there exist two degrees of freedom at each site.
Apart from the rotational and 
two translational modes of the entire plaquette, there are five normal modes 
described in \fig{fig:NormalModes}. 
These normal modes are, however, not independent
in the lattice and three constraints between the normal 
modes are necessary.\cite{AhnKH04nature}
%Thus only two normal modes are truly independent.
%and others can be represented in terms of them. 
Using the classical approximation,
we can write the harmonic part of the lattice Hamiltonian as follows.
\beq
H_\subbox{lat}^{0} = \f{1}{2}\sum_{i}\left\{
\sum_{n=1,2,3}\!\!\!A_n\left[e_n(\vi)\right]^2
\!\!+\!\!\!\!\sum_{m=+,-}\!\!\!\!B\left[s_m(\vi)\right]^2\right\},
\eeq
where it should be understood that only two of the normal modes 
are independent. 

It is easy to have a grip on the physical meaning of the parameters
in this Hamiltonian. The normal modes $s_\pm$ appear at
the domain walls so that $B$ is related to the domain
wall energy. It is also clear that $e_3$ mode
% is the mode that
couples directly
to the ENO; the domain boundaries of the lattice
distortions related to the $e_3$ modes will coincide with those of the 
ENO. 
%Here we implicitly assume that the elastic constants, especially $A_3$,
%are self-consistently determined after including the coupling to the
%nematic order.
It is expected that the $e_3$ mode is particularly affected
and will be soft. Thus we assume that $A_3$ is small while all the 
$A_n$ and $B$ are positive.  

Using the constraints, the lattice Hamiltonian with an anharmonic correction
can be rewritten as follows.
%only in terms of $e_1$ and $e_3$ modes as follows.
%Including an anharmonic correction, we obtain
\beq
%\nonumber
H_\subbox{lat}
= \sum_{\bq,a,b=1,3} 
e^*_a ({\bf q})\f{M_{ab}({\bf q})}{2} 
e_b ({\bf q})\!+\!\f{C_3}{4} \! \sum_i[e_3(\vec{i})]^4\!,
\eeq
with
%\!\!\!\!\!\!\mbox{where}
$M_{ab} =  \! A_a  \delta_{ab}\!+\! 
A_2 {\beta_a \beta_b \over \beta_2^2} + B {\beta_4 \over \beta^2_2} 
[ \beta_1 \delta_{ab} + \beta_3 (1-\delta_{ab})]$.
$C_3$ is a constant
%where 
%\beq
%M_{ab} = A_a \delta_{ab} + 
%A_2 {\beta_a \beta_b \over \beta_2^2} + B {\beta_4 \over \beta^2_2} 
%[ \beta_1 \delta_{ab} + \beta_3 (1-\delta_{ab})]
%\eeq
and
$\b_1(\bq) = 1 - \cos q_x \cos q_y$,
$\b_2(\bq) = - \sin q_x \sin q_y$,
$\b_3(\bq) = \cos q_x - \cos q_y$, and
$\b_4(\bq) = (1-\cos q_x)(1-\cos q_y)$.
%It turns out that it is necessary to include anharmonic terms to allow
%domain structures in the lattice distortions.
%The simplest way to 
%proceed is to include an anharmonic term in the $e_3$ mode. This is 
%because the domain boundaries of the nematic order would be strongly 
%coupled to those of the $e_3$ mode. Now the total lattice Hamiltonian
%is given by
%\beq
%H_\subbox{ph} = H_\subbox{ph}^\subbox{0}+ \f{C_3}{4} \sum_i [e_3({\vec i})]^4,
%\label{eqn:phonon-hamil}
%\eeq

{\bf Electron-lattice interaction}: We consider the Su-Schrieffer-Heeger type 
electron-lattice interaction,
\beq
H_\subbox{el-lat} =
g \sum_i  [ (d_{i+\hx}^x - d_i^x) c_{i+\hx}^\dag c_{i} + (x \leftrightarrow y) 
+\mbox{h.c} ],
\label{eqn:SSHmodel}
\eeq
where $d_{i}$ represents the lattice displacement.
% with the majority or minority spin
%that go through the nematic transitions. 
Writing this interaction in terms of the lattice normal modes, we get
\beq
H_\subbox{el-lat} = 
g\sum_{\bk,\bq}\left[
\f{f_{\bk0}e_1(\bq)+ f_{\bk1}e_3(\bq)}{(1+e^{iq_x})(1+e^{iq_y})}
c_{\bk+\bq}^\dag c_{\bk}
+\mbox{h.c}
\right],
\label{eqn:el-ph_hamil}
\eeq
where $f_{\bk0} = \cos k_x + \cos k_y$. Notice that the $e_3$ mode
couples to the ENO through the form factor $f_{\bk1}$
and the $e_1$ mode to the kinetic energy.

{\bf Domain structures}:
We numerically determine the electronic and lattice
configurations of the system described by
$H = H_\subbox{el}+H_\subbox{el-lat}+H_\subbox{lat}$.
The simplest domain wall solutions can be 
obtained by assuming that the spatial modulations
%associated with the domains
% of the ENO
exist only along the diagonal direction.
This makes the numerical works easier while one can 
still understand the key properties of the domain wall configurations.
We also assume that the $e_1$ mode is much stiffer than the
$e_3$ mode so that $e_1$ mode is not heavily affected by
the nematic transitions.
We will discuss later the role of the $e_1$ mode in the context
of the magnetostriction measurements in Sr$_3$Ru$_2$O$_7$.

The non-uniform ENO parameter and lattice configurations are 
determined by the self-consistent mean field theory described below. 
We compute the energy of the anharmonic part of the lattice Hamiltonian
in real space while the rest of the energy is computed in the reciprocal
space. We use the periodic boundary condition in the reciprocal space
and the fast-Fourier transform to connect the real and reciprocal space
computations. Here the number of discrete ${\bf q}$ points in the Brillouin
zone of the reciprocal space is related to the lattice size in the real space.
We first consider $H_{\rm el} + H_{\rm el-lat}$ and
determine the ENO parameter $Q (\bq;e_3)$ of the electronic state for 
a given initial lattice configuration $\{ e_3 \}$. The resulting electronic state 
$|\Phi (e_3) \rangle_{\rm el}$
is used to compute the energy $\tilde{E}_\subbox{el} (e_3)
= {}_{\rm el}\langle \Phi (e_3) | H_\subbox{el}+H_\subbox{el-lat}(e_3) 
|\Phi (e_3) \rangle_{\rm el}$.
Next we minimize the total energy 
$E_\subbox{total}(e_3) = \tilde{E}_\subbox{el}(e_3) + H_\subbox{lat}(e_3)$
by the Euler method\cite{AhnKH04nature}.
The resulting lattice configuration is in general different from the initial
one and it is then fed back to determine a new
%nematic order parameter or the
electronic state $| \Phi (e_3) \rangle$.
We update the lattice 
configuration $\{ e_3 \}$ following the maximum energy slopes with respect 
to $e_3(\bq)$ until $\f{\d E_\subbox{total}(e_3)}{\d e_3}=0$ is satisfied.
%We started from random initial configurations of the lattice and then determine
%the resulting self-consistent electronic and lattice configurations.

\begin{figure}[tbh]
\epsfxsize=6.0cm
\epsffile{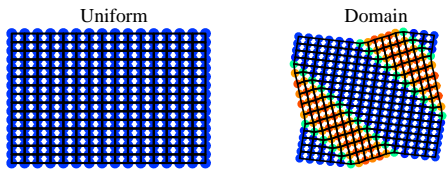}
\caption{
(Color online)
The red, blue, and green represent the negative, positive, and zero
ENO parameter respectively.
%The black lines denote the lattice structure. 
Here, $A_1 = 20$, $A_2 = 20$, $A_3 = 0.001$, $B=1$, $C_3 = 100$,
$t = 0.1$, $t'=0$, $F = 0.2$, and $g = 0.2$.
The electron filling is $\f{120}{256}$ per site to be slightly away from the
van Hove singularity.
\label{fig:Nematic_Lattice}
}
\end{figure}

The relative probability of getting the metastable domain solutions is obtained
by trying a large number (typically 50 in a given set) of random initial conditions. 
This probability is only $32\%$ for the $8 \times 8$ lattice, it increases
to $50 \%$ for the $18 \times 18$ lattice,
and
%it becomes
$63 \%$ for the $20 \times 20$ lattice
with 2$\sim$3 $\%$ error.
Thus the number of metastable domain solutions increases as
the system size increases in accord with similar computations in 
manganites.\cite{AhnKH04nature}  These domain solutions imply
a large number of local minima in the free energy and will be likely
chosen at finite temperatures by entropic effect. 
\fig{fig:Nematic_Lattice} shows a typical uniform and 
a meta-stable domain solution.
The qualitative features of these solutions do not depend on the
details of the parameters chosen here; for example, the electron filling
factor $\f{120}{256}$ with $t'=0$ is chosen just to avoid the van Hove 
singularity and a similar result is obtained for the half-filling with a
small finite $t'$.
%Notice that the domain walls of the lattice distortion coincide with 
%those of the nematic order. 

{\bf The stability of domains}:
\begin{figure}[tbh]
\epsfxsize=7cm
\epsffile{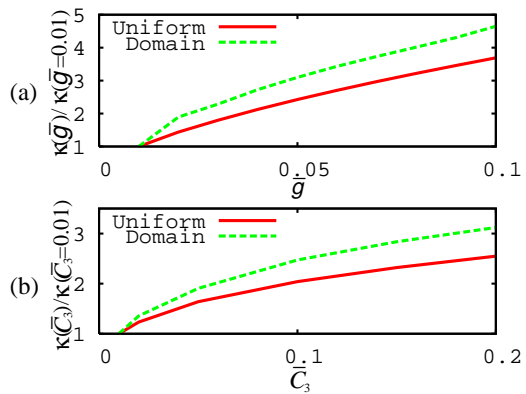}
\caption{
(Color online)
$\k$ as a function of 
(a) the electron-lattice coupling and (b) the anharmonicity.
We use $t'=0$, $F=2t$, and $B=20t$. 
$\bar{C}_3=0.05$ for (a) and $\bar{g} = 0.1$ for (b).
\label{fig:Laplacian}
}
\end{figure}
The stability of the domain configurations 
would be determined by the barrier potential 
in the energy landscape and the energy difference between the 
uniform state and the domain structure.

Here we consider an indirect criterion of the `local' stability,
$\kappa \equiv \sum_\bq
\f{\d^2}{\d e_3(\bq)^2}E_\subbox{total}$
evaluated at the local minima corresponding to the meta-stable
domain configurations. This quantity is the curvature of the energy 
topography in the hyperspace of lattice configurations. 
Larger $\kappa$ would correspond to a deeper local minimum, implying
perhaps a more stable configuration. Even though this quantity measures
the stability of the domains in the lattice deformations, the nematic domains
would follow because they are basically locked together.

Of particular interest is the stability with respect to the electron-lattice coupling 
$g$ and the strength of the anharmonicity $C_3$. 
\fig{fig:Laplacian} (a) and (b) show the dependence of $\kappa$ on 
the dimensionless coupling constants $\bar{g}={g \bar{e}_3}/{t}$
and $\bar{C}_3={C_3\bar{e}_3^4}/{t}$, where
$\bar{e}_3$ is the average of $e_3$ and about the order of $0.1$ lattice
constant. The increase of $\kappa$ in both of the uniform and domain 
solutions may lead to bigger barrier height and the faster increase of
$\kappa$ for the domain configuration may imply that
the domain structure at large $\bar{g}$ and $\bar{C}_3$ 
would be at a relatively deeper minimum. 

\begin{figure}[tbh]
\epsfxsize=7cm
\epsfysize=4cm
\epsffile{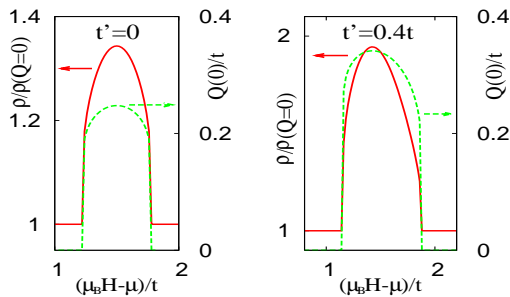}
\caption{
(Color online)
The dashed line denotes the uniform ENO parameter in the
mean-field theory.\cite{Khavkine04prb} The solid lines
represent the resistivity with $t'=0$, and $0.4t$. Here we use
$\rho/\rho (Q=0) = \f{\a}{\a- Q(0)^2}$ with $\a=0.24$ and $F = 1.974t$.
\label{fig:resistivity}
}
\end{figure}

{\bf Resistivity}:
%\section{Resistivity}
%\label{sec:Resistivity}
The large enhancement of the residual resistivity in 
the emergent phase may be explained by the scattering at 
the nematic domain walls.
We adopt the Landauer formula to estimate the conductance through
a domain wall, namely $G_\subbox{DW} = \f{e^2}{\pi\hbar}T$,
where $T$ is the transmission coefficient.\cite{Buttiker85prb}
%This transmittance coefficient arises from the Fermi velocity mismatch
%due to the different nematic phase between the adjacent domains.
%The Fermi velocity in nematic phase for a square lattice with
%next-nearest hopping can be written as
%$
%v_{Fx}^{(\pm)} = 
%\left(t\pm Q(0) - 2t'\cos k_y\right)\sin k_x$, and
%$v_{Fy}^{(\pm)} = \left(t\mp Q(0) - 2t'\cos k_x\right)\sin k_y$.
%\beqa
%v_{Fx}^{(\pm)} &=& 
%\left(t\pm Q(0) - 2t'\cos k_y\right)\sin k_x,~~~\mbox{and}  \\
%v_{Fy}^{(\pm)} &=& 
%\left(t\mp Q(0) - 2t'\cos k_x\right)\sin k_y.
%\eeqa
In the case of a domain wall parallel to the $b$-axis, the transmission coefficient 
along the $a$-axis (or the $x$-axis) is given by 
$ T = \sum_{\bk \in FS}T_\bk$,
where $T_\bk = 
v_{Fx}^{(+)}({\bf k}) v_{Fx}^{(-)} ({\bf k})/ 
|v_{Fx}^{(+)}({\bf k}) +v_{Fx}^{(-)} ({\bf k}) +iV_0|^2 $ and 
${\bf k} \in FS$ ensures that only the ${\bf k}$ vectors 
residing on the appropriate Fermi surface are included.
%This equation can be acquired by matching the boundary condition of
%plane waves of left and right region of nematic domain.
%\beq
%T = \f{v_{Fx}^{(+)}v_{Fx}^{(-)}}
%{\left|v_{Fx}^{(+)}+v_{Fx}^{(-)}+iV_0\right|^2}
%\eeq
Here ${\bf v}_{F}^{(\pm)}(\bk) = \frac{\partial}{\partial {\bf k}} 
\xi_\bk^{\subbox{ren},(\pm)} |_{{\bf k} \in FS}$, $V_0$ is the strength of
the scattering potential at the interface, and
$\pm$ sign denotes the direction of the nematic-ordered Fermi 
surfaces in the different sides of the domain wall.
%For $+(-)$ sign, the Fermi surface is elongated in $b$($a$)-direction.
%Since the electron density does not vary much at the nematic domain
%wall, we can assume the $V_0$ is very small.
If the scattering processes at different domain walls are incoherent,
the resistivity due to $N_\subbox{DW}$ domain walls would be
\beq
\rho_\subbox{DW} \propto N_\subbox{DW} G^{-1}_{DW}
= {N_\subbox{DW}}\left/\left( u_0 - u_1  Q^2 \right)\right.
\label{eqn:resistivity}
\eeq
%\beqa
%\nonumber
%G_\subbox{DW} &=& \f{e^2}{\pi\hbar}\f{1}{(2\pi)^2}
%\int\!dk_xdk_y
%\f{\left\{\left(t-2t'\cos k_y\right)^2+\D^2\right\}\sin^2 k_x}
%{4(t-2t'\cos k_y)^2\sin^2 k_x + V_0^2} \\
%&=&
%C_0 + C_1 \D^2
%\eeqa
where
$u_0 = \f{e^2}{\pi\hbar} \int_{\bk \in FS}\!
{d^2 k \over (2\pi)^2} 
\f{\left(t-2t'\cos k_y\right)^2\sin^2 k_x}
{4(t-2t'\cos k_y)^2\sin^2 k_x + V_0^2}$,
$u_1 = \f{e^2}{\pi\hbar} \int_{\bk \in FS}\!
{d^2 k \over (2\pi)^2} 
\f{\sin^2 k_x}
{4(t-2t'\cos k_y)^2\sin^2 k_x + V_0^2}$.
The resulting resistivity is shown in 
\fig{fig:resistivity} for $t'\not=0$ and $t'=0$ cases.
Notice that $t'\not=0$ case leads to an asymmetric profile
of the resistivity similar to that seen in the experiment.
The important point, however, is that the overall shape of the
resistivity follows the behavior of the ENO 
and this aspect is independent of the parameters.
%{\tt and reflects the behavior of the nematic order as a function of
%the external magnetic field.}

{\bf Summery and Discussion}:
The theory of ENO is consistent with 
several key experimental observations 
in Sr$_3$Ru$_2$O$_7$.\cite{Grigera04Science,Borzi07Science}
1) There exist two first order MMTs as the magnetic
field increases, 2) the electrons are itinerant in the entire phase diagram,
3) the emergent phase exists only in a window of magnetic fields.
4) the finite temperature transition to the emergent phase 
seems to be continuous.\cite{Grigera04Science,KeeHY05prb} 
In this paper, we have shown that this
theory combined with the coupling to the lattice leads to
nematic domain formation and this naturally explains
the high resistivity in the emergent phase.\cite{Grigera04Science,Borzi07Science} 
In particular, it is shown that the resistivity is determined by the magnetic field
dependence of the ENO parameter.

Some additional remarks are in order.
The magnetostriction data on Sr$_3$Ru$_2$O$_7$ 
show a sudden change in the $c$-axis lattice constant 
upon entering and exiting the emergent phase.\cite{Grigera04Science} 
This may be explained
by a relatively soft $e_1$ mode that is related to a change in the
$ab$-plane unit-cell area. Such a change would lead to elongation 
of the lattice along the $c$-axis to conserve the three dimensional unit-cell volume.
The jump in the kinetic energy of the electrons with the majority/minority
spin, via the coupling to the $e_1$ mode, would induce an abrupt 
change of the $c$-axis lattice constant at the nematic transitions. 

A very recent experiment 
observed an interesting dependence of the resistivity on the
relative orientation of the in-plane magnetic field and the current
direction.\cite{Borzi07Science} 
When they are parallel (perpendicular) to each other, 
the high resistivity disappears (is maintained).
This is reminiscent of the situation in the nematic quantum Hall 
states in high Landau levels.\cite{eisenstein}
A similar mechanism \cite{Jungwirth99prb} of the nematic
domain alignment by an in-plane magnetic field may work here as well.
This similarity, therefore, may make the relevance of the
nematic theory even stronger. It would be an excellent subject of future study.

We would like to thank Steve Kivelson, Andy Millis, and Andy MacKenzie for helpful
discussions. This work was supported by the NSERC of Canada,
the CRC, the CIAR, and KRF-2005-070-C00044.


\begin{thebibliography}{99}

\bibitem{Grigera01Science}
S. A. Grigera {\it et al.}, Science {\bf 294}, 329 (2001).

\bibitem{perry01}
R. S. Perry {\it et al.}, 
Phys. Rev. Lett. {\bf 86}, 2661 (2001).

\bibitem{Millis02prl}
A. J. Millis, {\it et al.},
Phys. Rev. Lett. {\bf 88}, 217204 (2002).

\bibitem{Green05prl}
A. G. Green, {\it et al.}, Phys. Rev. Lett. {\bf 95}, 086402 (2005).

\bibitem{Grigera04Science}
S. A. Grigera {\it et al.}, Science {\bf 306}, 1155 (2004). 

\bibitem{perry04}
R. S. Perry {\it et al.}, Phys. Rev. Lett. {\bf 92}, 166602 (2004). 

\bibitem{kitagawa}
K. Kitagawa {\it et al.}, Phys. Rev. Lett. {\bf 95}, 127001 (2005).

\bibitem{Binz06prl}
B. Binz {\it et al.}, Phys. Rev. Lett. {\bf 96}, 196406 (2006). 

\bibitem{KeeHY05prb}
H.-Y. Kee and Y. B. Kim, Phys. Rev. B {\bf 71}, 184402 (2005).

\bibitem{Borzi07Science}
R. A. Borzi, {\it et al.}, Science {\bf 315}, 214 (2007)
%R. A. Borzi, S. A. Grigera, J. Farrell, R. S. Perry, S. Lister, S. L. Lee,  D. A. Tennant, 
%Y. Maeno, and A. P. Mackenzie, 
%``Formation of a nematic fluid at high fields in Sr$_3$Ru$_2$O$_7$,
%to be published in Science.
%preprint, submitted to Science (2006).    

\bibitem{Oganesyan01prb}
V. Oganesyan, S. A. Kivelson, and E. Fradkin, Phys. Rev. B {\bf 64}, 195109 (2001).

\bibitem{Khavkine04prb}
I. Khavkine {\it et al.}, Phys. Rev. B {\bf 70}, 155110 (2004). 

\bibitem{AhnKH04nature}
K. H. Ahn, T. Lookman, and A. R. Bishop, Nature {\bf 428}, 401 (2004); 
K. H. Ahn {\it et al.}, Phys. Rev. B {\bf 68}, 092101 (2003), and references therein. 

\bibitem{Buttiker85prb}
M. B\"uttiker {\it et al.}, Phys. Rev. B {\bf 31}, 6207 (1985). 

\bibitem{eisenstein}
M. P. Lilly {\it et al.}, Phys. Rev. Lett. {\bf 82}, 394  (1999); 
Phys. Rev. Lett. {\bf 83}, 824 (1999).

\bibitem{Jungwirth99prb}
T. Jungwirth {\it et al.}, Phys. Rev. B {\bf 60}, 15574 (1999).


\end{thebibliography}
\end{document}